\documentclass[noshowpacs, twocolumn,notitlepage,preprintnumbers,secnumarabic,amsmath,superscriptaddress, showkeys, amssymb,nofootinbib,longbibliography]{revtex4-1}
\usepackage{bm}           
\usepackage{graphicx,bm}
\usepackage{caption}
\usepackage{natbib}
\usepackage{subfloat}
\usepackage{subcaption}
\usepackage[font=small,labelfont=bf]{caption}
\usepackage{gensymb}
\usepackage{float}
\usepackage[pdftex,bookmarks,colorlinks,breaklinks]{hyperref}  
\usepackage{mathrsfs}
\usepackage{color}
\usepackage[table]{xcolor}
\usepackage{amsmath}
\usepackage{natbib}
\usepackage{aas_macros}

\makeatletter

    \def\CT@@do@color{%
      \global\let\CT@do@color\relax
            \@tempdima\wd\z@
            \advance\@tempdima\@tempdimb
            \advance\@tempdima\@tempdimc
    \advance\@tempdimb\tabcolsep
    \advance\@tempdimc\tabcolsep
    \advance\@tempdima2\tabcolsep
            \kern-\@tempdimb
            \leaders\vrule
                    \hskip\@tempdima\@plus  1fill
            \kern-\@tempdimc
            \hskip-\wd\z@ \@plus -1fill }
    \makeatother

\begin{document}
\title{Technical Considerations in Magnetic Analogue Models}
\author{Patrick W. M. Adams}
\email{pw.adams@uct.ac.za} \affiliation{Astrophysics, Cosmology and Gravity Centre, Department of Mathematics and Applied Mathematics, University of Cape Town, Rondebosch 7701, Cape Town, South Africa}
\author{Bob Osano} 
\email{bob.osano@uct.ac.za} \affiliation{Astrophysics, Cosmology and Gravity Centre, Department of Mathematics and Applied Mathematics, University of Cape Town, Rondebosch 7701, Cape Town, South Africa}
\affiliation{Academic Development Programme, Science, Centre for Higher Education Development, University of Cape Town, Rondebosch 7701, Cape Town, South Africa}

\begin{abstract}
The analogy between vorticity and magnetic fields has been a subject of interest to researchers for a considerable period of time, mainly because of the structural similarities between the systems of equations that govern the evolution of the two fields. We recently presented the analysis of magnetic fields and hydrodynamics vorticity fields \cite{osanoadams2016-1, osanoadams2016-2} and argued for a formal theory of {\it analogue magnetism}. This article provides in depth technical details of the relevant considerations for the simulation procedures and extends the analyses to a range of fluids.
\end{abstract} 


\date{\today}

\maketitle

\section{Introduction}

For many years scientific enquiry could only be pursued by theoretical analysis or by direct experimentation. Simulation has emerged in recent years as powerful, although indirect, new mode of enquiry. In particular, the advent of computers and computational methods is allowing researchers to make progress in the study of problems in Fluid Dynamics, Magnetohydrodynamics (MHD), Turbulence and Astrophysics, thereby leading to better understanding of the properties of the substance being analysed. With the improvement in technology over the years and the coming-of-age of high-performance computing, many problems in these areas that were previously deemed intractable have now become tractable, with much time being spend on developing efficient computer codes for specialised applications in these fields.\\ \par Some of these codes are specially designed to run on a variety of computer architectures, ranging from a single machine perhaps possessing a multi-core CPU or several GPUs, to larger, more complicated architectures which are the case for cluster- or supercomputers which typically possess thousands of GPU units and CPU cores. Many of these codes are freely available, for example the popular ZEUS code which was presented in a series of test problems considered in a series of papers by Stone \textit{et al} \cite{stone1992a, stone1992b, stone1992c}, and the \textsc{Pencil Code}, which is used to obtain the results presented in this paper and whose design are rationale are described in \cite{brandenburg2001, brandenburgdobler2002, pencilman}. With the advent of general-purpose computing on graphics processing units (GPGPU), many codes similar to ZEUS and \textsc{Pencil} such as GPUPEGAS, which is used to simulate interacting galaxies through MHD \cite{igor2013}, FARGO3D \cite{benitezmasset2016}, which is used mainly for applications  planetary-disk interactions in forming stellar systems, and RAMSES-GPU \cite{kestener2010} which was ported to the GPU from the popular RAMSES MHD code \cite{teyssier2002}.

Hydromagnetic turbulence is an example of an area that has benefitted greatly from advances in computing. The study of turbulence by Kolmogorov \cite{kolmogorov1991a, kolmogorov1991b, kolmogorov1991c} and others and its application in turbulent dynamo theory is another. The gradients of the various slopes observed in the energy spectrum of a magnetic field amplified by a turbulent dynamo has been the subject of study for many authors (e.g. \cite{haugenea2004, brandenburgdobler2002, brandenburg2001, brandenburg2005rev, haugen2003} and others). In particular, in \cite{haugenea2004}, discussed the importance of a fine enough mesh resolution in order to correctly capture important features in the magnetic energy spectrum. Though we do not consider turbulence in this paper, the discussion of mesh resolution is key when assessing the quality of solutions obtained so as to avoid any arti-facts that are solely due to a mesh that is possibly too coarse. 

The search for a theory of {\it analogue magnetism}, that links Fluid Dynamics to Electromagnetism, is entering a new phase that stands to benefit from strides  already made in computing and computer technology. Such a theory will allow the indirect study of many more complicated aspects of Electromagnetism than is currently possible.  Today, however, such a theory is rudimentary and references made to any possible physical analogy between the two fields are only made in passing. In this paper, we aim to further discuss the analogy between the magnetic field present in a charged fluid, and the vorticity field present in a neutral fluid with an emphasis on technical aspects of simulations. A specific case of this has already been discussed in a prior work \cite{osanoadams2016-1}, where we examined the analogy between the magnetic and vorticity field where the magnetic diffusion and kinetic viscosity were kept equal, and where source terms in the relevant equations were neglected. This paper extends the work in \cite{osanoadams2016-1} and examines a bigger range of diffusion and viscosity values; in particular, we do not demand that these two values be equal. Physically, this is akin to testing for an analogy between magnetic field and vorticity in different types of fluids. The case involving the consideration of source terms in the relevant equations is explored elsewhere \cite{osanoadams2016-2}.

The idea of a possible analogy between Electromagnetism and Fluid Dynamics is not new, in fact Maxwell himself noticed similarities between the magnetic vector potential $\mathbf{A}$ and the velocity field $\mathbf{u}$, describing the flow of a fluid in which magnetic field lines are embedded \cite{maxwell1861, siegel2002}. Indeed, an analogy is even drawn between the Biermann battery term in the two-fluid Induction Equations and the baroclinic term in the evolution equations for fluid vorticity \cite{brandenburg2005rev, kulsrud1996}. In recent times, many authors have gone further and have attempted to analyse similarities in physical behaviours between these two fields \cite{arbab2011, kulsrud1996, martins2009, martins2012}, which is a key step in developing the theory of analogy. In particular, \cite{kulsrud1996} discusses the analogy between the magnetic field and vorticity, noting that their root mean square (rms) strengths appear to saturate around the same time, as well as with the same strength in the case for a turbulent dynamo acting on the magnetic field.
\par This work considers more technical aspects of the simulations. In the following sections, we first introduce the governing equations of MHD in the forms in which they will be solved for our purposes, and spend some time discussing the analogous system consisting of the Induction and Vorticity equations. A brief description of the \textsc{Pencil Code} is then given and its solution strategy is discussed. We then turn to discussing the issue of mesh resolution selection before presenting our main results which focus on the conditions under which an analogy between the magnetic field and vorticity would hold. For our purposes, we discuss the results obtained in terms of the magnetic Prandtl number, $\mathrm{Pr_M}$, particular to that run in order to ascertain the conditions under which the analogy will hold, given known values of magnetic diffusivity and kinematic viscosity. The behaviour observed in the results of our simulations is then discussed in relation to physical behaviour that has already been reported in literature thus far.

\section{Governing Equations}
In this section we collect all equations relevant to the two systems we intend to analyse and compare.

\subsection{The Magnetohydrodynamics Equations} 
The hydrodynamics of a charged fluid are governed by the equations of Magnetohydrodynamics (MHD). These include the fluid continuity equation, the Navier-Stokes equations and the magnetic Induction Equations. These are listed below in units of $\mu_0=1$, where $\mu_0$ is the permiability of free space.

\begin{subequations}
	\begin{align}
		\frac{\mathrm{D}\ln\rho}{\mathrm{D}t} & = -\nabla\cdot\mathbf{u} \label{eq:cont} \\
		\frac{\mathrm{D}\mathbf{u}}{\mathrm{D}t} & = -\frac{\nabla p}{\rho} + \nu\nabla^2\mathbf{u} \label{eq:ns} \\
		\frac{\partial\mathbf{A}}{\partial t} & = \mathbf{u}\times\mathbf{B} + \eta\nabla\times\mathbf{B}. \label{eq:ind}
	\end{align}
\end{subequations}

Here, $\rho$ is the fluid density, $\mathbf{u}$ is the veloctiy field, $p$ is the fluid pressure, $\mathbf{A}$ the magnetic vector potential, $\mathbf{B}$ the magnetic flux density (related to $\mathbf{A}$ via $\mathbf{B} = \nabla\times\mathbf{A}$ and hereafter simply referred to as the magnetic field), $\nu$ the fluid viscosity and $\eta$ the magnetic diffusivity. Equations (\ref{eq:ind}) are traditionally written in terms of and solved for $\mathbf{B}$; writing them in terms of and solving for $\mathbf{A}$ automatically fulfills the divergenceless condition on $\mathbf{B}$. This set of equations would also normally include an evolution equation for the internal energy, $U$, or entropy, $s$, of the system; however, we do not consider the effects of entropy in this paper. Equations (\ref{eq:cont}) --- (\ref{eq:ind}) in their given forms are solved numerically in the results presented. We now consider the relevant analogous equations that are the primary theme in this paper.

\subsection{The Relevant Analogous System}
The equations we consider are the following:
\begin{subequations}
	\begin{align}
		\frac{\partial\boldmath{\omega}}{\partial t} & = \nabla\times(\mathbf{u}\times\boldmath{\omega}) + \nu\nabla^2\boldmath{\omega} \label{eq:vort} \\
		\frac{\partial\mathbf{B}}{\partial t} & = \nabla\times(\mathbf{u}\times\mathbf{B}) + \eta\nabla^2\mathbf{B} \label{eq:indb}.
	\end{align}
\end{subequations} 

It can be shown that equations (\ref{eq:indb}) and (\ref{eq:vort}) are simply the curled versions of equations (\ref{eq:ns}) and (\ref{eq:ind}) given above. The form of eqn (\ref{eq:vort}) demands that $\mathbf{u}$ is necessarily solenoidal. As with the case for the magnetic vector potential, $\mathbf{A}$, $\mathbf{u}$ may be recovered similarly from $\boldmath{\omega}$ through integration. Written in this form, it is clear that the two equations are structurally identical. When a Biermann battery term is added to eqn (\ref{eq:indb}), which arises naturally when considering a fluid consisting of both protons and electrons, the system takes the form:
\begin{subequations}

	\begin{align}
		\frac{\partial\boldmath{\omega}}{\partial t} & = \nabla\times(\mathbf{u}\times\boldmath{\omega}) + \nu\nabla^2\boldmath{\omega} - \frac{\nabla p_{\mathrm{(nf)}}\times\nabla\rho_{\mathrm{(nf)}}}{\rho^2_{\mathrm{(nf)}}} \label{eq:vortbatt} \\
		\frac{\partial\mathcal{B}}{\partial t} & = \nabla\times(\mathbf{u}\times\mathcal{B}) + \eta\nabla^2\mathcal{B} + \frac{\nabla p_{\mathrm{(cf)}}\times\nabla\rho_{\mathrm{(cf)}}}{\rho^2_{\mathrm{(cf)}}} \label{eq:indbatt},
	\end{align}
\end{subequations}
where the speed of light in vacuo $c$ has been set to 1 in eqn (\ref{eq:indbatt}). Using the rescaling for $\mathbf{B}$ in \cite{osanoadams2016-2}, and where the subscripts (nf) and (cf) denote a neutral and charged fluid respectively. Note that the appearance of the corresponding ``battery term'' in eqn (\ref{eq:vortbatt}) is due to the presence of the pressure term in the Navier-Stokes equations (eqns (\ref{eq:ns})). Eqns (\ref{eq:vortbatt}) and (\ref{eq:indbatt}) are still structurally identical and may be written in a unified form as \cite{osanoadams2016-2}:
\begin{equation}
\label{eq:uni}
\frac{\partial \mathcal{D}}{\partial t} = \nabla\times(\mathbf{u}\times\mathcal{D}) + d\nabla^2\mathcal{D} - \frac{\nabla p_{\mathrm{(\star f)}}\times\nabla\rho_{\mathrm{(\star f)}}}{\rho_{\mathrm{(\star f)}}^2},
\end{equation} 

where $\mathcal{D} = \boldsymbol{\omega}$ or $-\mathcal{B}$, $d = \nu$ or $\eta$ and ($\star$f) refers to either a neutral or charged fluid. Though we do not consider the effects of the Biermann battery term in this paper, we shall still compare the root mean square (rms) strengths of $\boldsymbol{\omega}$ and $-\mathbf{B}$ when presenting and discussing the results of our simulations in the following section.

\section{The Pencil Code}

The \textsc{Pencil Code}\footnote{\url{http://pencil-code.nordita.org}} is a publicly-available high-order finite differences MHD code that is well-suited to the study of various problems in MHD, including forced MHD turbulence, studies of dynamos, weakly-compressible flows and convection problems \cite{pencilman}, it may be extended to study limits of MHD approximations and even predictions resulting from new extensions to Maxwells equations such as in \cite{osano2016}, and its accompanying MHD extension \cite{genmaxwmhd}. Many MHD codes often make use of spectral schemes in solving the equations of MHD due to their high accuracy, but are often limited to only a small subsection of MHD applications due to their complexity. High-order finite difference codes are becoming increasingly popular due to their ease of implementation, parallelization and suitability for the study  of a larger number of MHD problems. 
\par As mentioned before, the \textsc{Pencil Code} comes from the latter family of MHD codes. The code itself is modular, in the sense that different variables and physical processes are consolidated into modules which can be included or excluded from any particular run as the need arises \cite{pencilman}. As an example: it is possible to simulate both MHD and normal hydrodynamics by including or excluding the code's \texttt{Magnetic} module which what we have done in this study. The \textsc{Pencil Code} solves eqns (\ref{eq:cont}) -- (\ref{eq:ind}) in their given forms. 
\par The \textsc{Pencil Code} is itself a non-conservative code, solving the the MHD equations in their so-called non-conservative form \cite{brandenburgdobler2002, pencilman}. Thus, the quality of solutions can be checked by monitoring how well certain conserved quantities (such as entropy, mass and momentum) are conserved for the duration of a partiular run. These conserved quantities are thus only conserved up to the discretisation error of the numerical scheme, rather than up to machine accuracy \cite{pencilman}. For more details on the code's technical aspects, the reader is referred to \cite{brandenburg2001, pencilman}.
\\
\section{Simulation Results and Discussion} 
 We now present the results of our simulations.
\subsection{Selecting Mesh Size}
We first present the results of initial simulations in which we compare the different mesh size  and their suitability for our purposes, after which we present the main results. For these simulations, we consider a neutral fluid which evolved according to a simplified system consisting only of eqns (\ref{eq:cont}) and (\ref{eq:ns}) and having different values of $\nu$. Simulation boxes of sizes $32^3$, $64^3$ and $128^3$ are used and the temporal evolution of $\mathbf{u}_\mathrm{rms}$ and $\boldsymbol{\omega}_\mathrm{rms}$ observed. A summary of simulation parameters are listed in table \ref{tab:simpars1}.
\begin{table}[H]
	\centering
	{\renewcommand{\arraystretch}{1.5}%
		\begin{tabular}{c|c|c|c}
			\hline \hline
			$\nu$ & $32^3$ & $64^3$ & $128^3$ \\
 			\hline 
			 \rowcolor{gray}$0$ & Run $1$ & Run $5$ & Run $9$ \\
			$10^{-5}$ & Run $2$ & Run $6$ & Run $10$ \\
			\rowcolor{gray}$10^{-3}$ & Run $3$ & Run $7$ & Run $11$ \\
			$10^{-1}$ & Run $4$ & Run $8$ & Run $12$ \\
			\hline \hline
		\end{tabular}}
	\caption{{\it The simulation parameters used to obtain the results for mesh selection. Rows in grey indicate run results presented in this section.}}
	\label{tab:simpars1}
\end{table} The run results are in general similar, and hence we only present those of runs 1 and 3 in this section for illustration. 

\begin{widetext}
~
\begin{figure}
  \centering
     \caption{Simulation results for mesh selection runs with velocity and vorticity equations with $\nu=0$ and $\nu=10^{-3}$. }
    		\begin{subfigure}[b]{0.350\textwidth}
       		\includegraphics[width=\textwidth]{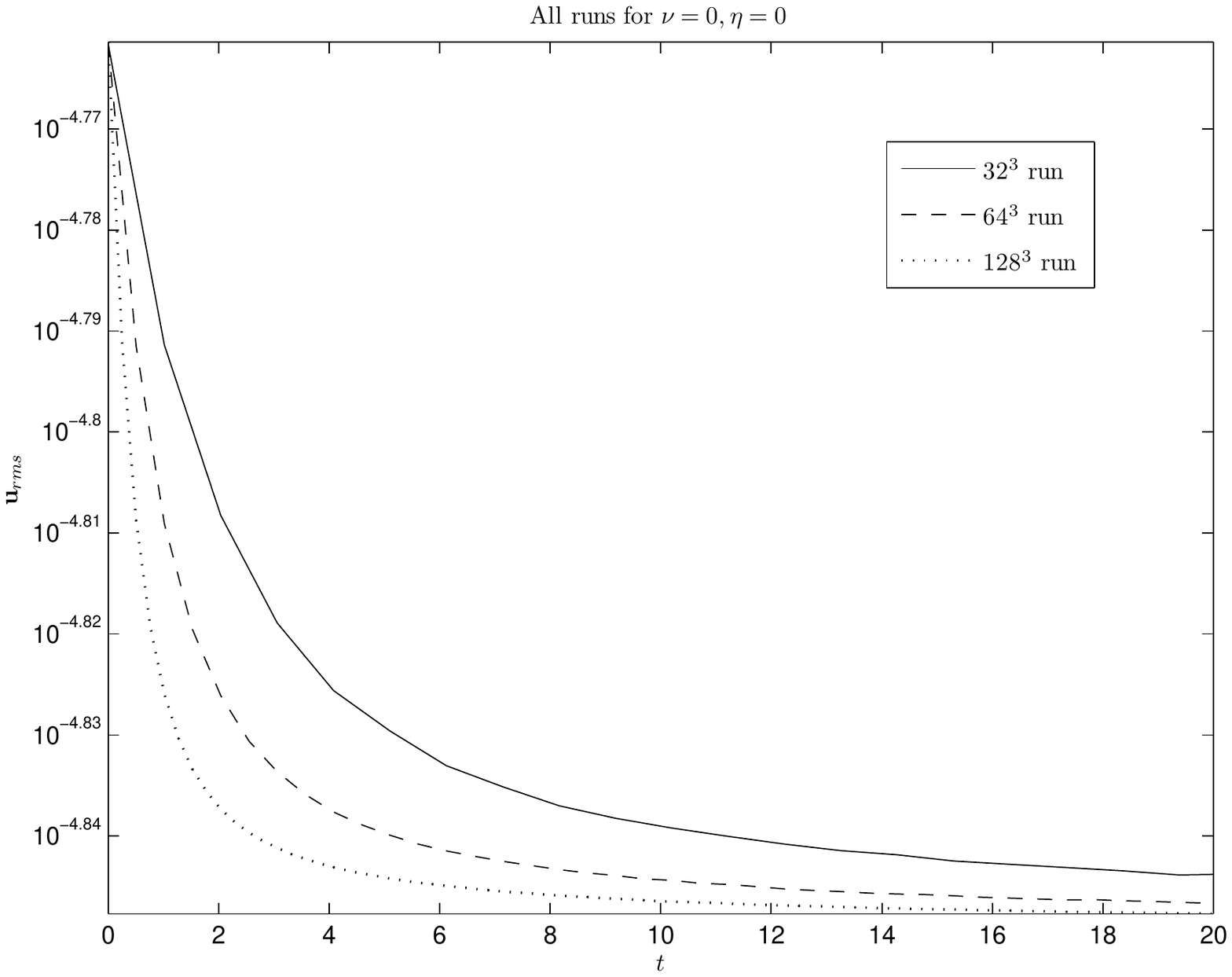}
       			 \subcaption{\it{ The first figure gives the plots of $\mathbf{u}_\mathrm{rms}$ versus time for all the $32^3$, $64^3$ and $128^3$ boxes. Results for all boxes show a monotonically decaying solution, with the decay becoming gradually more pronounced for a finer mesh. }}
   					 \label{fig:unutestrun0}
\end{subfigure} \qquad 
\begin{subfigure}[b]{0.350\textwidth}
      \includegraphics[width=\textwidth]{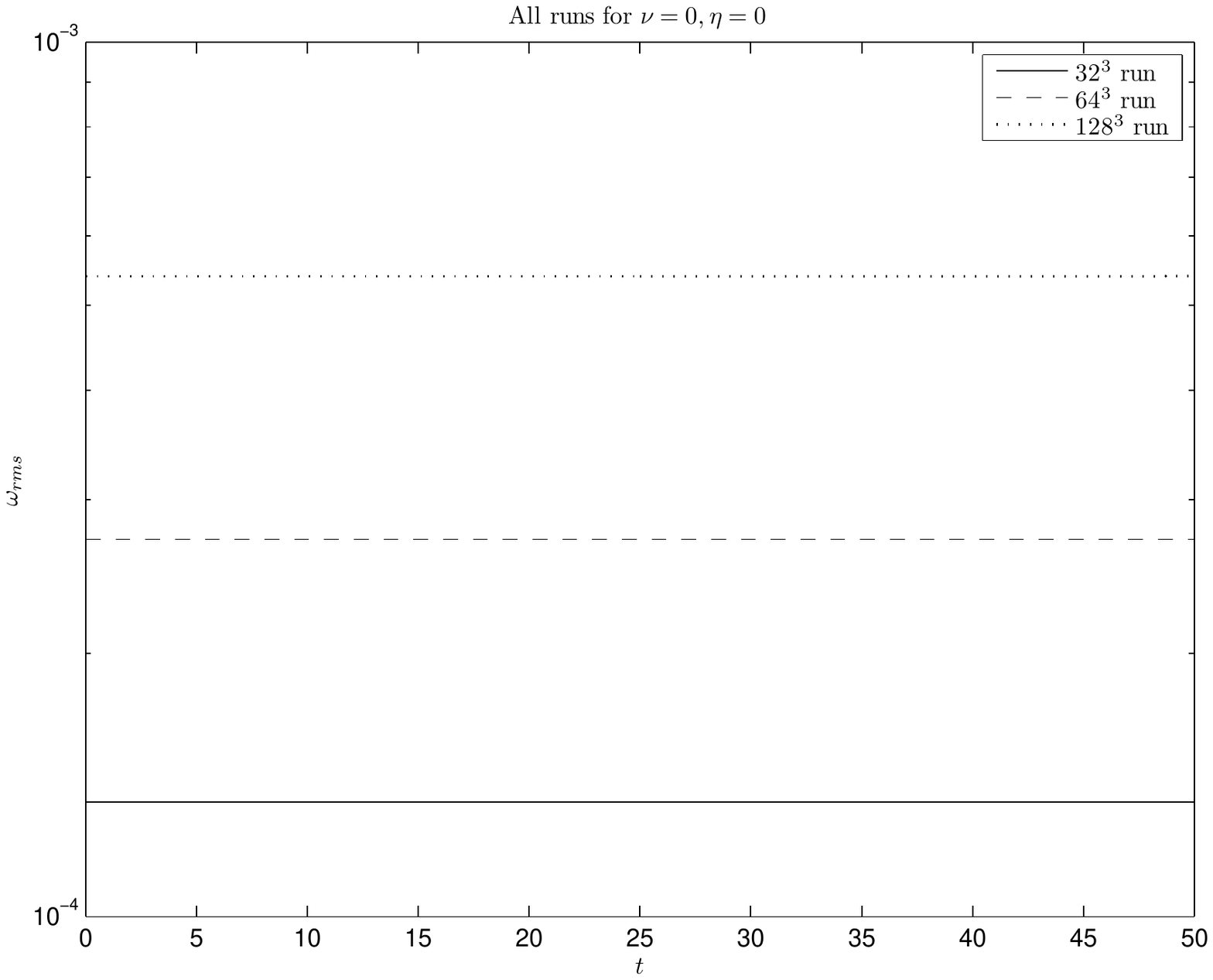}
       	\subcaption{\it The plots of $\boldsymbol{\omega}_\mathrm{rms}$ versus time for all the $32^3$, $64^3$ and $128^3$ boxes. Due to the very slow decay of $\mathbf{u}_\mathrm{rms}$, there is evidently no change in $\boldsymbol{\omega}_\mathrm{rms}$ for all the meshes considered. Due to the $128^3$ box producing the $\mathbf{u}_\mathrm{rms}$ solution with the most pronounced decay, the corresponding $\boldsymbol{\omega}_\mathrm{rms}$ solution appears to be the strongest.}
        		\label{fig:onutestrun0}
 \end{subfigure}\\
     \begin{subfigure}[b]{0.40\textwidth}
        \includegraphics[width=\textwidth]{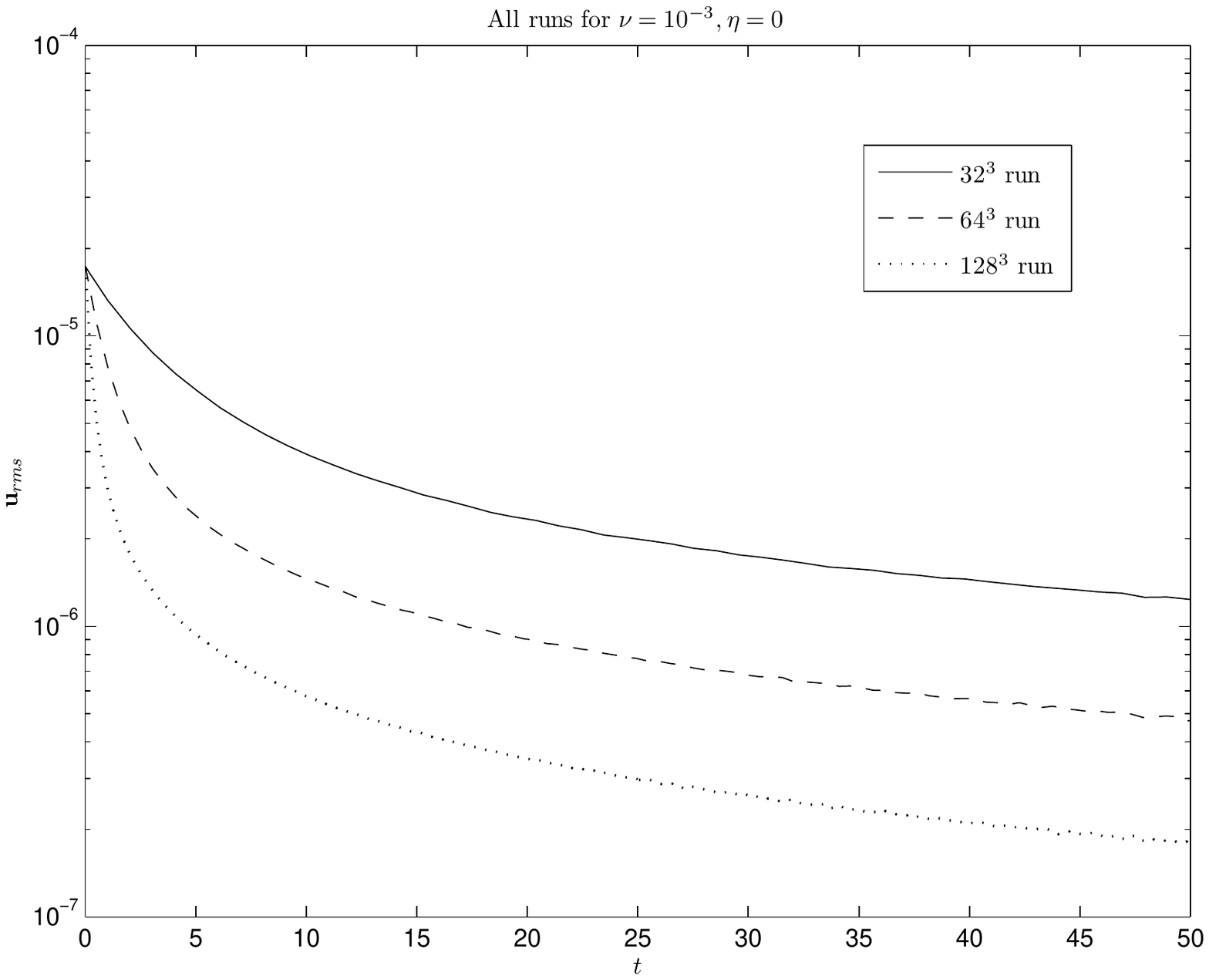}
         \subcaption{\it he plots of $\mathbf{u}_\mathrm{rms}$ versus time for all the $32^3$, $64^3$ and $128^3$ boxes. A monotonically decaying solution is again observed, with the decay becoming gradually more pronounced for a finer mesh.}
             \label{fig:unutestrun-3}
    \end{subfigure}~\qquad
        \begin{subfigure}[b]{0.40\textwidth}
        \includegraphics[width=\textwidth]{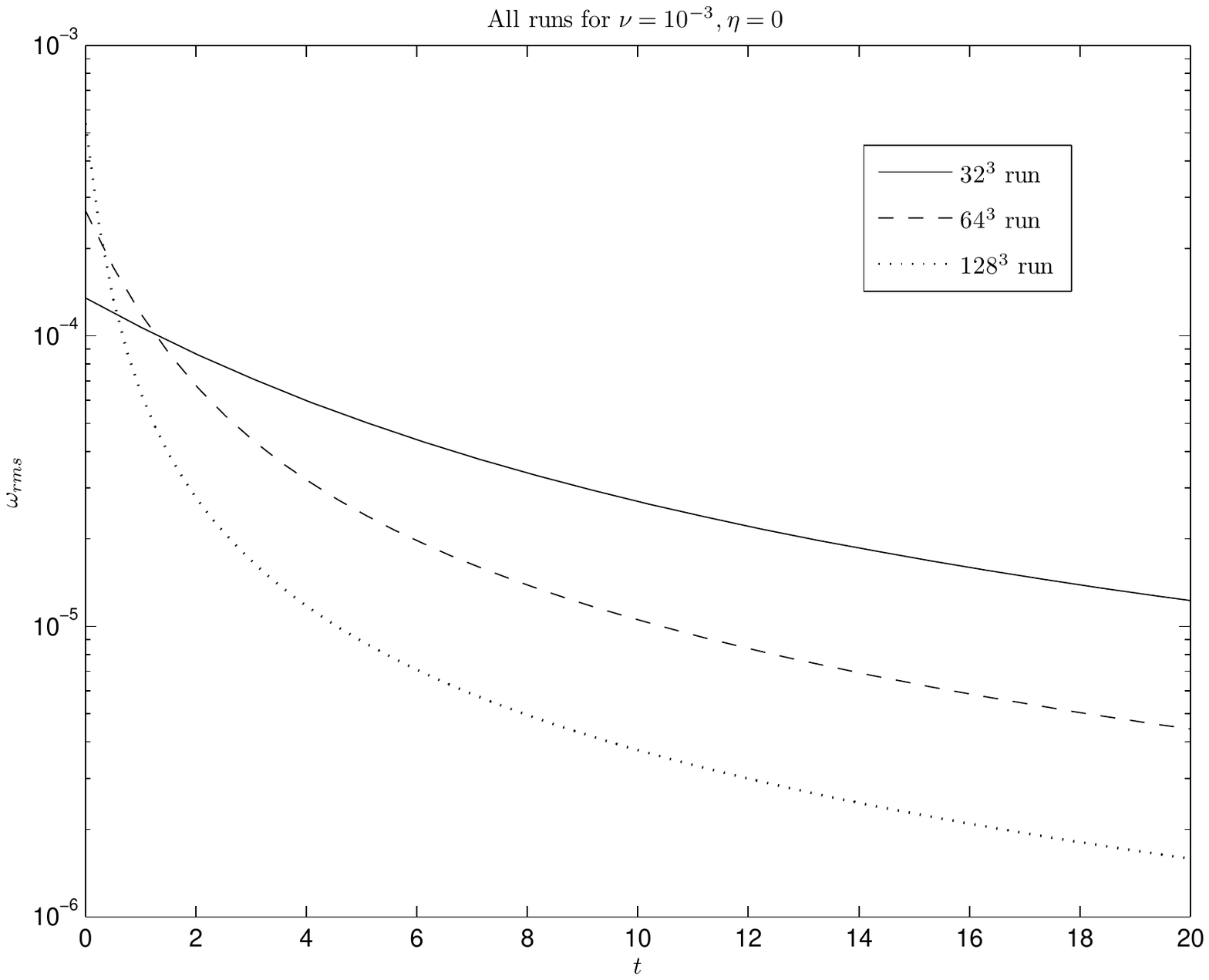}
        \subcaption{\it The plots of $\boldsymbol{\omega}_\mathrm{rms}$ versus time for all the $32^3$, $64^3$ and $128^3$ boxes. Monotonic decay in $\boldsymbol{\omega}_\mathrm{rms}$ is now present due to the effects of the diffusive term present in the evolution equations. Once more, the more pronounced decay of the $128^3$ solution results in an initially stronger$\boldsymbol{\omega}_\mathrm{rms}$ which then decays quickly.}
       \label{fig:onutestrun-3}
     \end{subfigure}\\
  \end{figure}\end{widetext}
  In run 1, we set $\nu = 0$, corresponding to an inviscid fluid (thereby eliminating the diffusive term in eqn (\ref{eq:ns})). The evolutions of $\mathbf{u}_\mathrm{rms}$ and $\boldsymbol{\omega}_\mathrm{rms}$ are presented in figures \ref{fig:unutestrun0} and \ref{fig:onutestrun0}. All three of the considered meshes indicate that $\mathbf{u}_\mathrm{rms}$ experiences marginal decay, which could be approximated using a straight flat line ( consistent with a constant terrm). The magnitude of $\boldsymbol{\omega}_\mathrm{rms}$ remains constant for the duration of the run. It should be noted here that the $128^3$ box which produced a slightly weaker $\mathbf{u}_\mathrm{rms}$ produces the strongest $\boldsymbol{\omega}_\mathrm{rms}$ and vice versa for the case of the $32^3$ box. This is to be expected due to the fact that the $128^3$ box's $\mathbf{u}_\mathrm{rms}$ solution shows the most pronounced decay over the duration of the run, thus producing a stronger $\boldsymbol{\omega}_\mathrm{rms}$.

For run 3 we set set $\nu = 10^{-3}$, corresponding to a viscous fluid where the corresponding eqn (\ref{eq:ns}) has a non-vanishing diffusive term. We observe more pronounced decay patterns for $\mathbf{u}_\mathrm{rms}$ for all the boxes under consideration, the effect of non vanishing diffusion. Unlike in the previous case, we also note that $\boldsymbol{\omega}_\mathrm{rms}$ is now also decaying due to the presence of a non-vanishing diffusive term in its evolution equation. For the same reasons given previously, the $128^3$ box produces the strongest initial $\boldsymbol{\omega}_\mathrm{rms}$ which then decays sharply, whilst the $32^3$ box produces the weakest initial $\boldsymbol{\omega}_\mathrm{rms}$ which then decays slowly when compared to the other two boxes.

Disregarding the differences in strengths of $\mathbf{u}_\mathrm{rms}$ and $\boldsymbol{\omega}_\mathrm{rms}$ produced by the three boxes under consideration, we note that the results obtained are qualitatively identical. Due to fineness of mesh, computation times for the $64^3$ and $128^3$ boxes are also very much longer than compared to the $32^3$ box; the former two stopping at an earlier value of the simulation time $t$ when compared to the $32^3$ box. In addition, the nature of the simulations conduced in this paper do not require very fine mesh. For these reasons, we choose to use the $32^3$ box as our preferred mesh for simulation. We now turn to main results of the simulations.

\subsection{Main Results}
As mentioned above, a $32^3$ periodic box of dimensions $2\pi\times2\pi\times2\pi$ was utilized for these simulations. All initial conditions were set to Gaussian noise of small amplitude, and the temporal growth of the rms strengths of the relevant quantities (i.e. $\mathbf{u}_\mathrm{rms}$, $\boldsymbol{\omega}_\mathrm{rms}$, $\mathbf{A}_\mathrm{rms}$ and $\mathbf{B}_\mathrm{rms}$) observed and compared.

The values of $\eta$ and $\nu$ were adjusted in each run in such a way that three specific cases of the magnetic Prandtl number (defined as $\mathrm{Pr_M} = \nu/\eta$) are investigated: $\mathrm{Pr_M}\ll1$, $\mathrm{Pr_M} = 1$ and $\mathrm{Pr_M}\gg1$. The special cases of $\mathrm{Pr_M} = 0$ and $\mathrm{Pr_M} \rightarrow \infty$ were also investigated. Of particular interest to us are the cases where $\mathrm{Pr_M}\neq1$, where the analogy between the evolution equations for $\mathbf{u}$ and $\mathbf{A}$, and thus also for $\boldmath{\omega}$ and $\mathbf{B}$, breaks down. Simulation parameters together with the associated magnetic Reynolds number, $\mathrm{Re_M} \equiv u_\mathrm{rms}L/\eta$, for that run are summarized in table \ref{tab:simpars2}. 

\begin{table}[H]
	\centering
	{\renewcommand{\arraystretch}{1.5}%
		\begin{tabular}{c|c|c|c|c}
			\hline \hline
			Run & $\nu$ & $\eta$ & $\mathrm{Pr_M}$ & $\mathrm{Re_M}$ \\
 			\hline 
			13 & $0$ & $10^{-5}$ & $0$ & 10.9033 \\
		 	14 & $10^{-3}$ & $10^{-1}$ & $10^{-2}$ & $6.7317\times10^{-5}$ \\
			15 & $10^{-3}$ & $10^{-3}$ & $1$ & 0.0031 \\
			16 & $10^{-3}$ & $10^{-5}$ & $10^2$ & 0.3063 \\
			17 & $10^{-5}$ & $0$ & $\infty$ & $\infty$ \\
			\hline \hline
		\end{tabular}}
	\caption{Summary of the simulation parameters together with the associated magnetic Reynolds number for that run used to obtain the results presented in this paper.}
	\label{tab:simpars2}
\end{table} In all of the simulations conducted, the aim was to determine under which conditions the analogy between the magnetic field and vorticity field would hold true. As suggested in our previous work \cite{osanoadams2016-1} as well as \cite{kulsrud1996}, the analogy should hold true for the case where $\mathrm{Pr_M} = 1$; though the other cases are also of interest in observing the behaviour of the fluids under consideration. 

\subsection{Evolution of the velocity and vorticity fields}
\begin{widetext}
~
 \begin{figure}
 	\centering
	 \caption{Simulation results for $\mathbf{u}_\mathrm{rms}$, $\boldsymbol{\omega}_\mathrm{rms}$, $\mathbf{A}_\mathrm{rms}$ and $\mathbf{B}_\mathrm{rms}$ versus time our main simulation runs. }
         \begin{subfigure}[b]{0.38\textwidth}
        		\includegraphics[width=\textwidth]{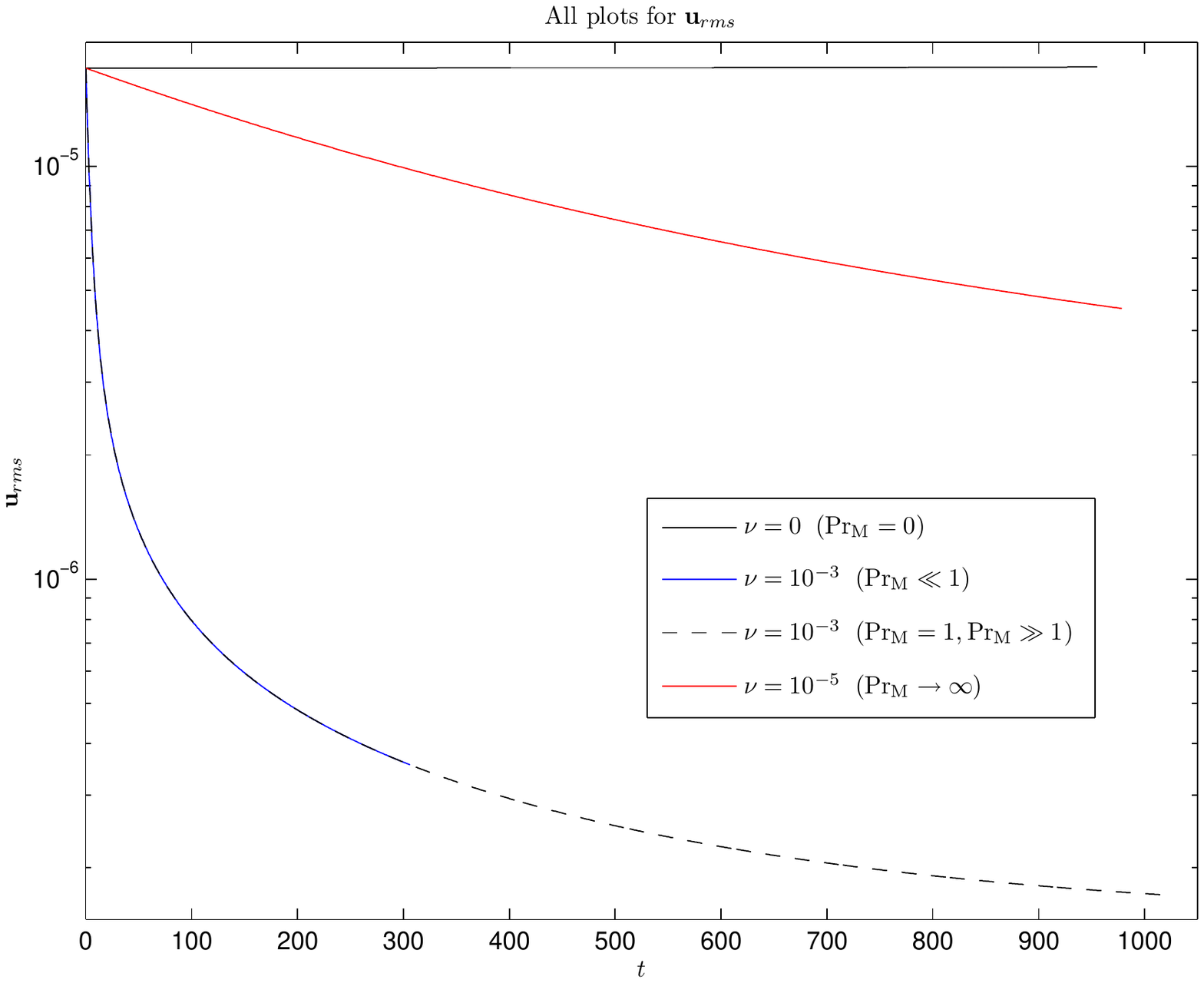}
        			\subcaption{\it The plots of $\mathbf{u}_\mathrm{rms}$ versus time for all values of $\mathrm{Pr_M}$. It is clear that for strong dissipation, the rms strength decays. Magnetic Prandtl numbers are displayed in order to connect the relevant $\mathbf{u}_\mathrm{rms}$ solution to its corresponding $\mathbf{A}_\mathrm{rms}$ counterpart(s).}
     			 \label{fig:uvst}
    	\end{subfigure}~\qquad\qquad\qquad	
	\begin{subfigure}[b]{0.38\textwidth}
       		 \includegraphics[width=\textwidth]{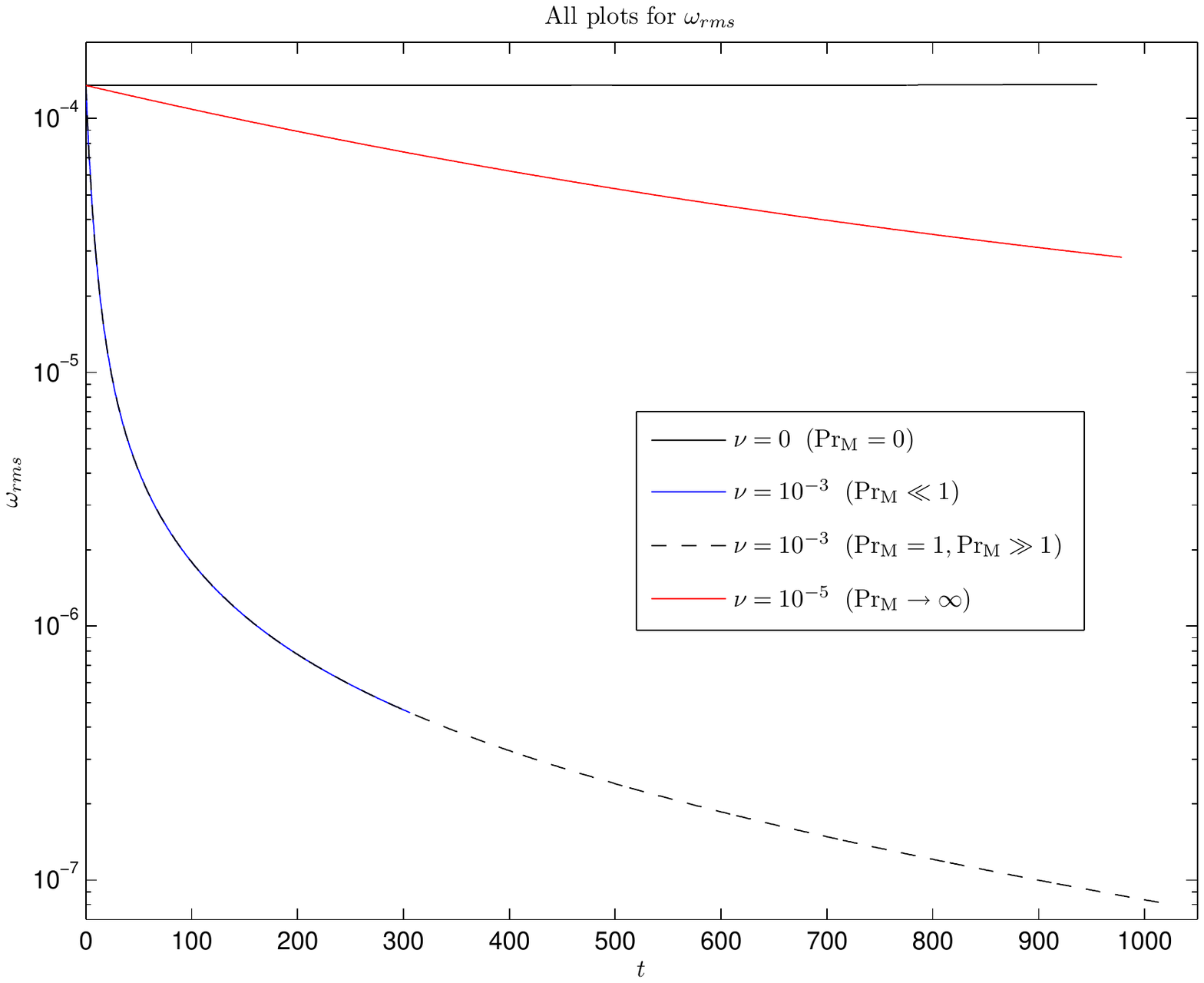}
        		\subcaption{\it The plots of $\boldsymbol{\omega}_\mathrm{rms}$ versus time for all values of $\mathrm{Pr_M}$. It is clear once more that for strong dissipation, the rms strength decays. As before, magnetic Prandtl numbers are displayed in order to connect the relevant $\boldsymbol{\omega}_\mathrm{rms}$ solution to its corresponding $\mathbf{B}_\mathrm{rms}$ counterpart(s).}
     			 \label{fig:ovst}
        \end{subfigure}\\
        \begin{subfigure}[b]{0.380\textwidth}
       		 \includegraphics[width=\textwidth]{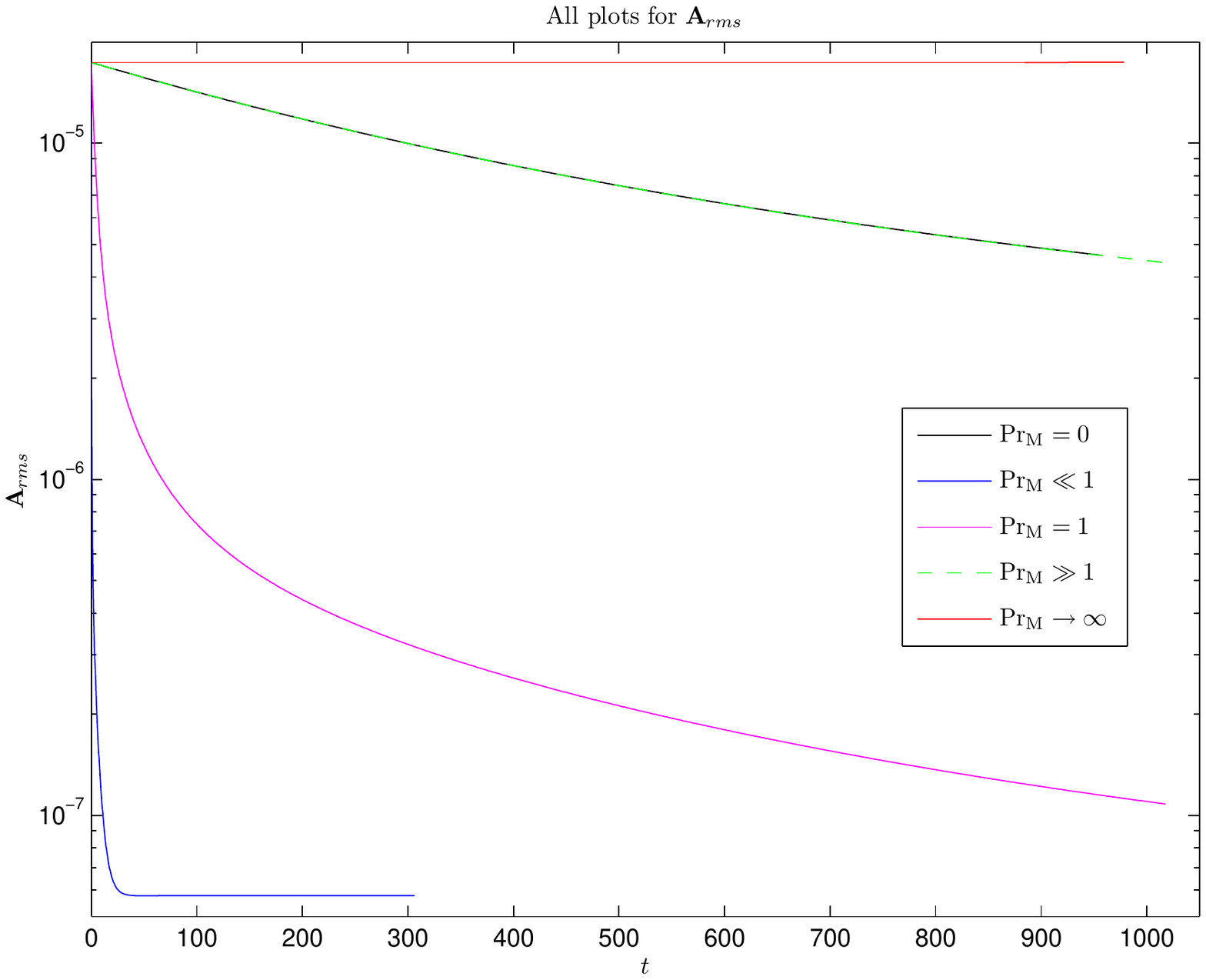}
        		\caption{\it The plots of $\mathbf{A}_\mathrm{rms}$ versus time for all values of $\mathrm{Pr_M}$. It is clear that for strong dissipation, the rms strength decays. Due to the runs for $\mathrm{Pr_M}=0$ and $\mathrm{Pr_M}\gg1$ having little difference between them, their lines (the green and black dashed) appear superimposed on each other.}
     			 \label{fig:avst}
        \end{subfigure}~\qquad\qquad\qquad
        \begin{subfigure}[b]{0.380\textwidth}
       		 \includegraphics[width=\textwidth]{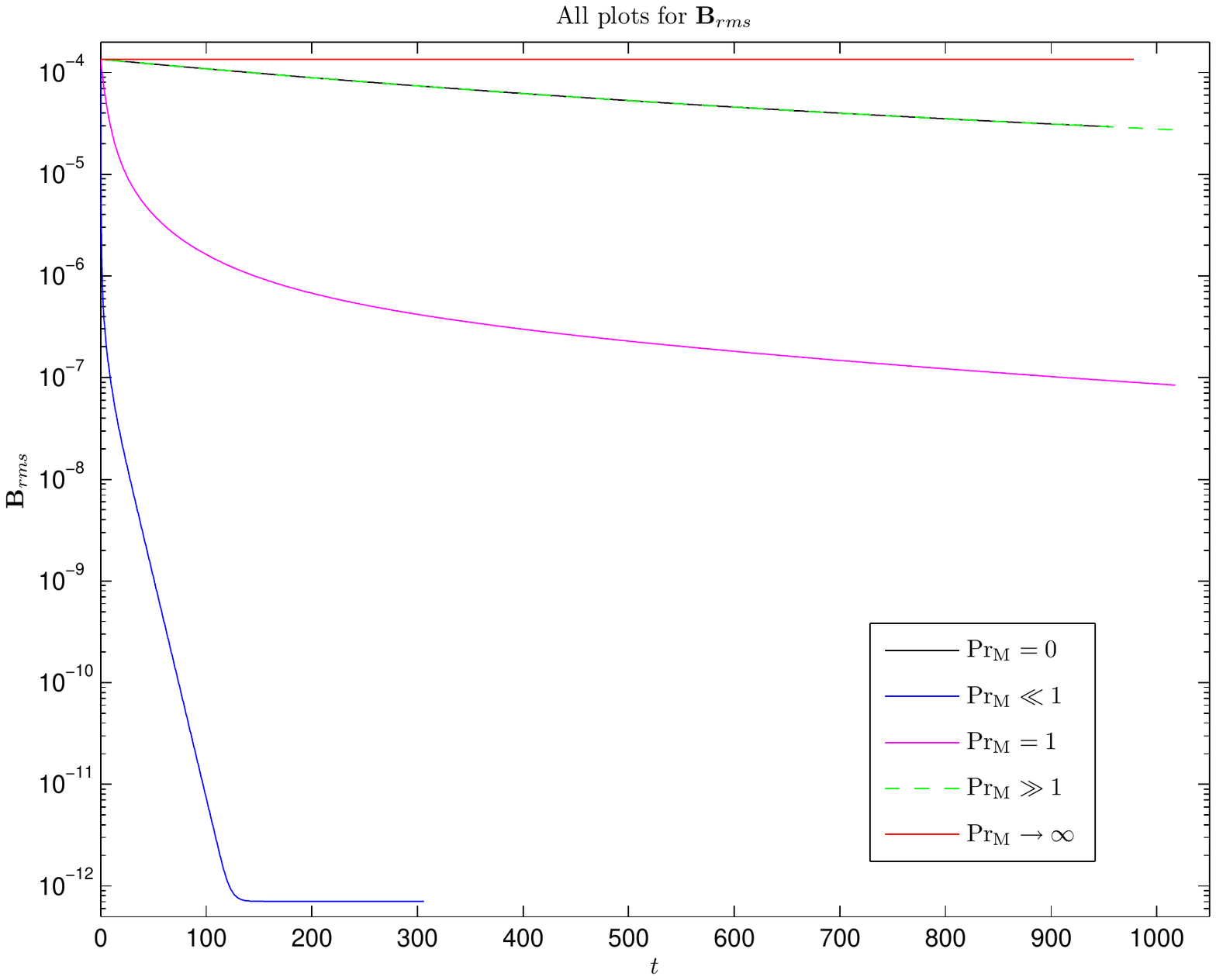}
        		\caption{\it The plots of $\mathbf{B}_\mathrm{rms}$ versus time for all values of $\mathrm{Pr_M}$. It is clear that for strong dissipation, the rms strength decays. Due to the runs for $\mathrm{Pr_M}=0$ and $\mathrm{Pr_M}\gg1$ having little difference between them, their lines (the green and black dashed) appear superimposed on each other.}
     			 \label{fig:bvst}
        \end{subfigure}\\
   \end{figure}
\end{widetext}

We first consider the temporal evolution of $\mathbf{u}_\mathrm{rms}$ and $\boldsymbol{\omega}_\mathrm{rms}$ with respect to different values of $\nu$. In particular, we are interested in conditions that we would lead to growing field strengths for both quantities. For simplicity, the Lorentz Force term was dropped from the Navier-Stokes Equations, meaning that there is essentially no way in which the magnetic field can interact with the fluid. Physically, we are then considering a system in which the back-reaction by the magnetic field on the fluid plays is essentially negligible. In light of this it is more sensible to discuss the temporal evolution of $\mathbf{u}_\mathrm{rms}$ (and hence $\boldsymbol{\omega}_\mathrm{rms}$) with respect to different values of $\nu$. 

Note, however, that the fluid naturally interacts with the magnetic field via the induction term in eqns (\ref{eq:ind}). Cases involving the consideration of the back-reaction and other non-linear effects will be considered in an upcoming work \cite{turbonset}.

Examining the temporal evolution behaviour displayed by the velocity and vorticity fields in figures \ref{fig:uvst} and \ref{fig:ovst}, it is clear that as $\nu\rightarrow0$, the exponential decay in the rms strengths observed becomes visibly slower until eventually the value of $\nu=0$ is reached, producing exponentially-growing rms field strengths. From the point-of-view of eqns (\ref{eq:ns}), this behaviour is expected, as $\nu$ essentially amplifies the effect of the diffusive term in a particular way.

As have chosen to examine the case of eqns (\ref{eq:indbatt}) without the Biermann battery term, effectively also dropping the pressure term from eqns (\ref{eq:ns}), and thus the corresponding "battery term'' from eqns (\ref{eq:vortbatt}), we also observe no accelerated decay in $\mathbf{u}_\mathrm{rms}$ and $\boldsymbol{\omega}_\mathrm{rms}$ which may be attributed to the presence of these terms.

\subsection{Evolution of the magnetic vector potential and magnetic flux density}

Results of simulating $\mathbf{A}_\mathrm{rms}$ and $\mathbf{B}_\mathrm{rms}$ for different values of $\mathrm{Pr_M}$ are displayed in figures \ref{fig:avst} and \ref{fig:bvst}. From the temporal evolution behaviour of both $\mathbf{A}_\mathrm{rms}$ and $\mathbf{B}_\mathrm{rms}$, it appears that as $\mathrm{Pr_M}$ increases, the rate of the exponential decay observed in the rms strengths for both fields becomes progressively slower until a turning point is reached beyond which the fields begin to grow.

The cases for $\mathrm{Pr_M}=0$ and $\mathrm{Pr_M}\gg1$ are particularly interesting: initially the rms strengths of $\mathbf{A}_\mathrm{rms}$ and $\mathbf{B}_\mathrm{rms}$ for the case of $\mathrm{Pr_M}\gg1$ are somewhat greater than the rms strengths of these quantities for the case of $\mathrm{Pr_M}=0$. However, after a finite amount of time (around 97s--98s for $\mathbf{A}_\mathrm{rms}$ and 154s -- 155s for $\mathbf{B}_\mathrm{rms}$) there is a cross-over and $\mathbf{A}_\mathrm{rms}$ and $\mathbf{B}_\mathrm{rms}$ for the case of $\mathrm{Pr_M}=0$ end up being somewhat stronger for the remainder of the run. This cross-over is not seen in figures \ref{fig:avst} and \ref{fig:bvst} due to the fact that the difference between the strengths of $\mathbf{A}_\mathrm{rms}$ and $\mathbf{B}_\mathrm{rms}$ for these fields is very small; the lines describing their temporal evolution thus appear to be superimposed on each other. 

 This crossover and final difference in strength can explain the apparent contradiction by recalling that we had observed exponentially-growing field strengths for $\mathbf{u}_\mathrm{rms}$ and $\boldsymbol{\omega}_\mathrm{rms}$ for the case of $\nu=0$ (and hence $\mathrm{Pr_M}=0$). As the velocity field interacts with the magnetic field, we would expect to see a slightly stronger magnetic field rms strength for the case of $\mathrm{Pr_M}=0$ after some time, as the velocity field rms strength is growing exponentially, despite the exponential decay in rms strength of the former field.

 \begin{widetext}
~
 \begin{figure}
 	\centering
	 \caption{Simulation results for $\mathbf{u}_\mathrm{rms}$ versus $-\mathbf{A}_\mathrm{rms}$ and $\boldsymbol{\omega}_\mathrm{rms}$ versus $-\mathbf{B}_\mathrm{rms}$ for our main simulation runs.}
         \begin{subfigure}[b]{0.40\textwidth}
        		\includegraphics[width=\textwidth]{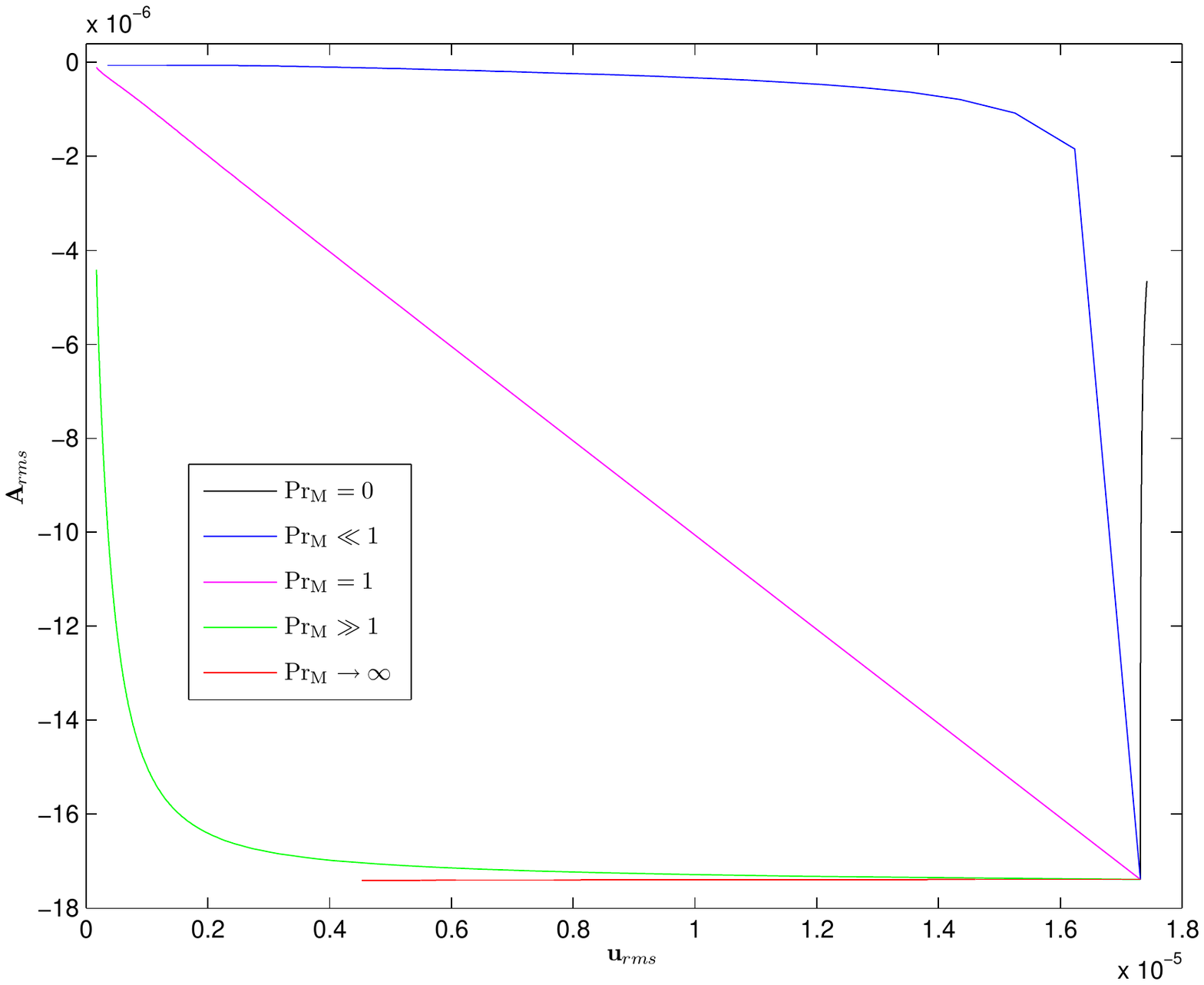}
        			\caption{\it $\mathbf{u}_\mathrm{rms}$ versus $-\mathbf{A}_\mathrm{rms}$ graphically. The special cases of $\mathrm{Pr_M}=0$ and $\mathrm{Pr_M}\rightarrow\infty$ appear to form an "envelope'' around the cases of $\mathrm{Pr_M}=1$ and $\mathrm{Pr_M}\neq1$; it is apparent that for $\mathrm{Pr_M}=1$, the analogy holds exactly.}
     			 \label{fig:uvsa}
    	\end{subfigure}~\qquad\qquad\qquad
	\begin{subfigure}[b]{0.40\textwidth}
       		 \includegraphics[width=\textwidth]{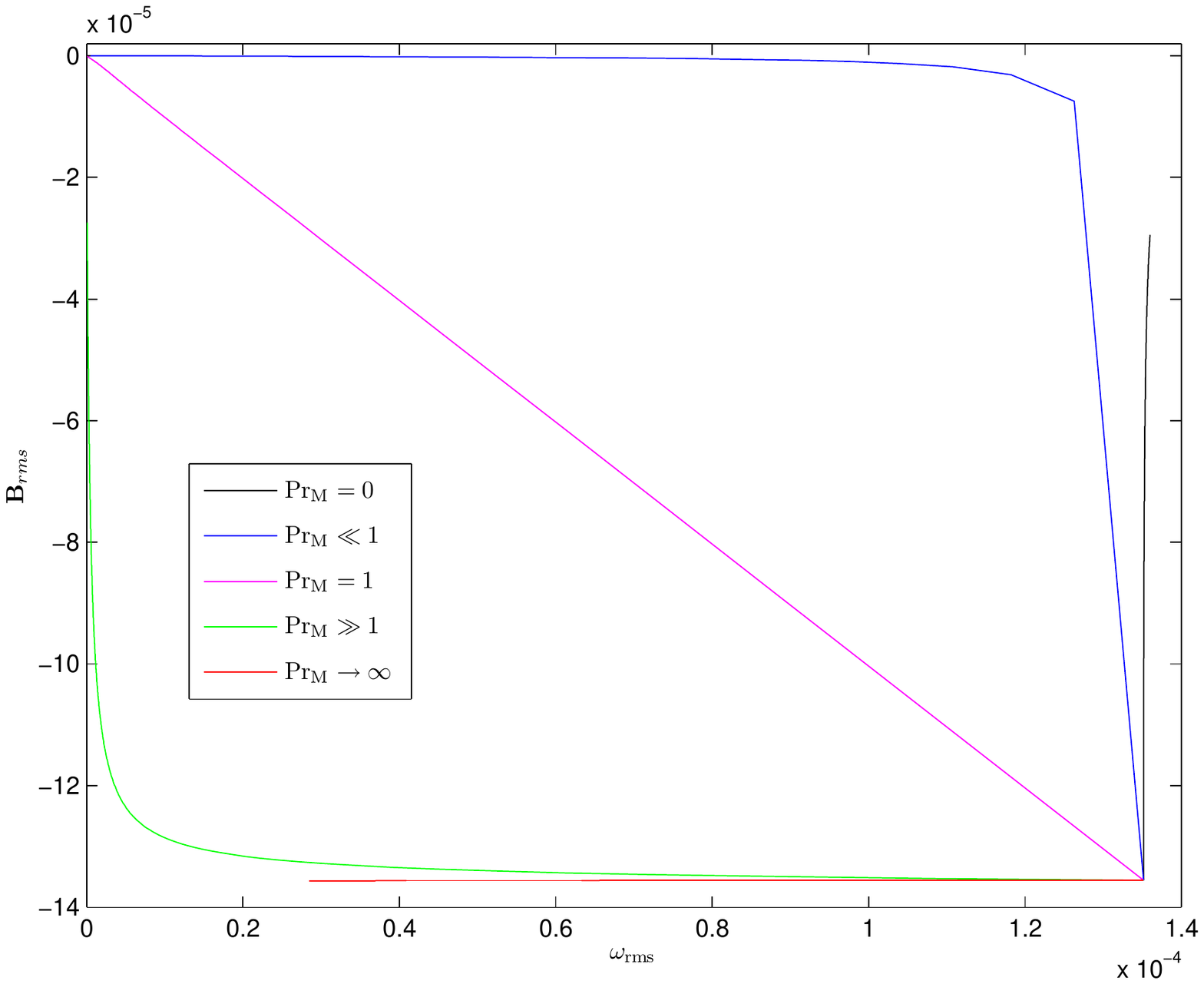}
        		\caption{\it\it $\boldsymbol{\omega}_\mathrm{rms}$ versus $-\mathbf{B}_\mathrm{rms}$ graphically. The special cases of $\mathrm{Pr_M}=0$ and $\mathrm{Pr_M}\rightarrow\infty$ again appear to form an "envelope'' around the cases of $\mathrm{Pr_M} = 1$ and $\mathrm{Pr_M}\neq1$. Once more it is apparent that the analogy holds exactly for $\mathrm{Pr_M}=1$.}
     			 \label{fig:ovsb}
        \end{subfigure}\\

\begin{subfigure}[b]{0.40\textwidth}
       		 \includegraphics[width=\textwidth]{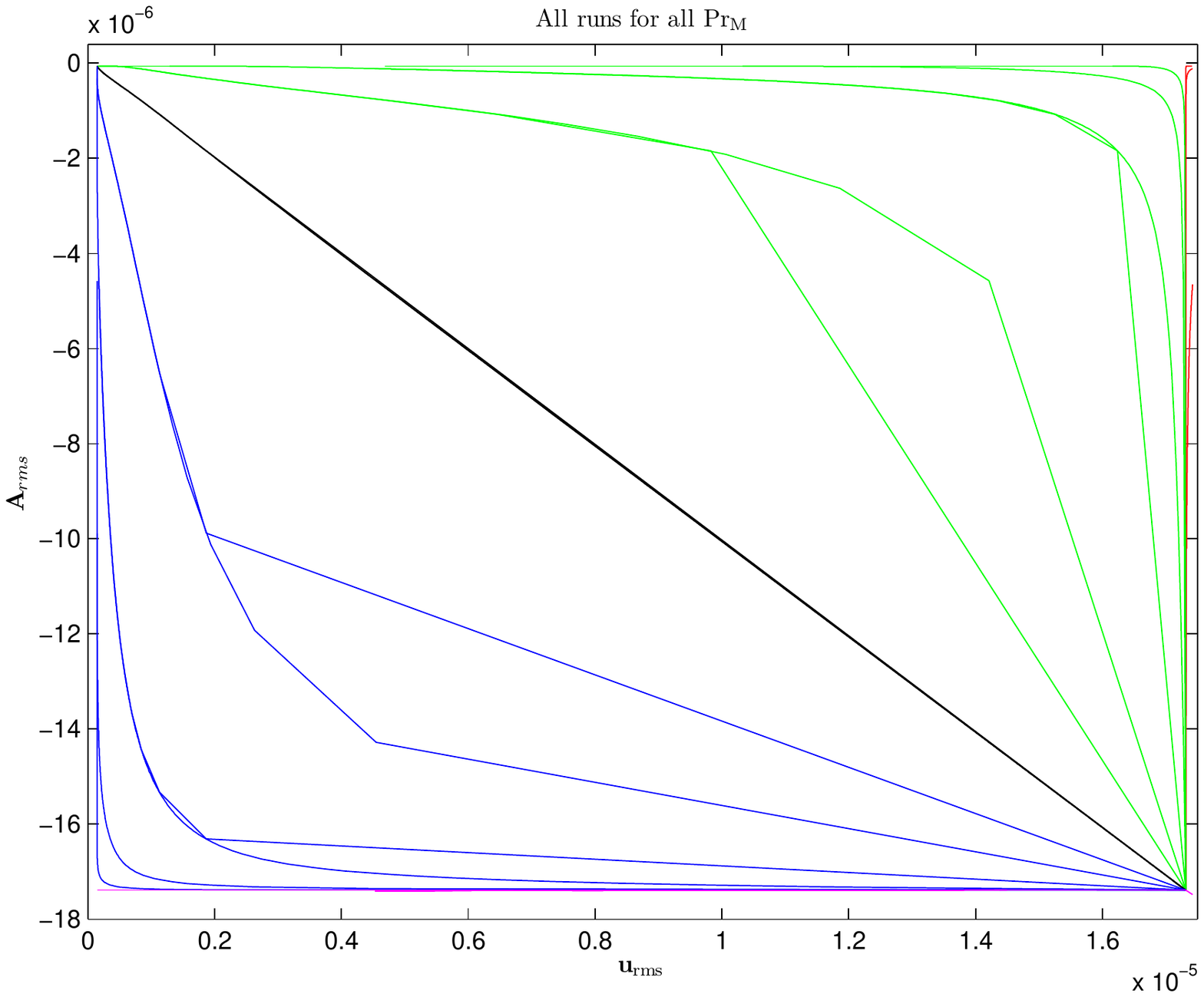}
        		\caption{\it $\mathbf{u}_\mathrm{rms}$ versus $-\mathbf{A}_\mathrm{rms}$ graphically for all runs considered. It is clear that the analogous relationship between $\mathbf{u}_\mathrm{rms}$ versus $\mathbf{A}_\mathrm{rms}$ holds exactly for the cases of $\mathrm{Pr_M}=1$.}
     			 \label{fig:uvsaallprm}
        \end{subfigure}~\qquad\qquad\qquad
        \begin{subfigure}[b]{0.40\textwidth}
       		 \includegraphics[width=\textwidth]{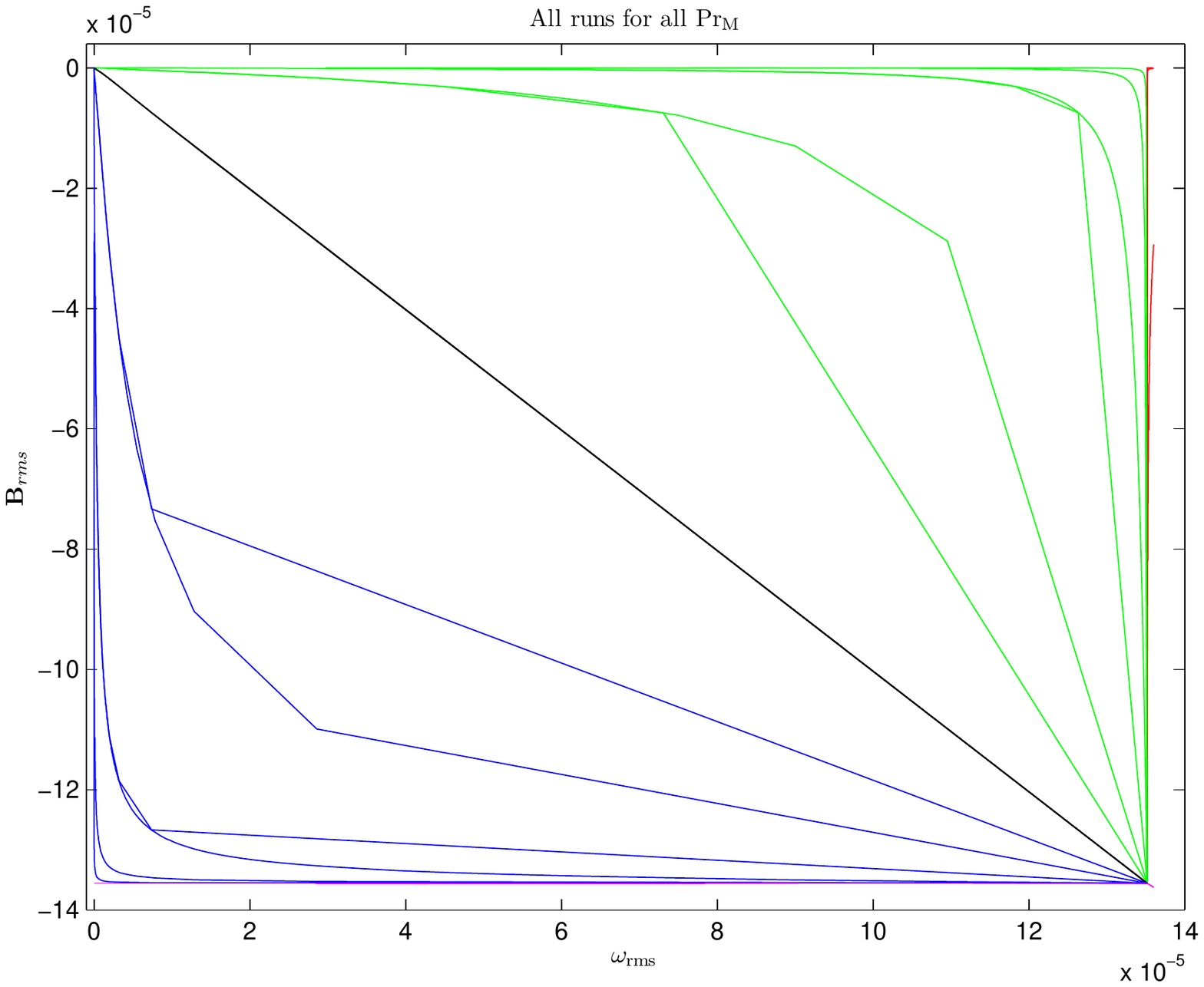}
        		\caption{\it $\boldsymbol{\omega}_\mathrm{rms}$ versus $-\mathbf{B}_\mathrm{rms}$ graphically. Again, it is clear that the analogous relationship between $\boldsymbol{\omega}_\mathrm{rms}$ versus $\mathbf{B}_\mathrm{rms}$ holds exactly for the cases of $\mathrm{Pr_M}=1$.}
     			 \label{fig:ovsballprm}
        \end{subfigure}\\
   \end{figure}
\end{widetext}

\subsection{The analogy between the vorticity and magnetic fields}
We now turn to discussing the analogy between eqns (\ref{eq:vort}) and (\ref{eq:indb}), as well as their counterparts, eqns (\ref{eq:ns}) and (\ref{eq:ind}). The simulation results discussed in this section are presented in figures \ref{fig:uvsa} and \ref{fig:ovsb}.

As was noted in our previous work \cite{osanoadams2016-2}, the analogy between the vorticity and magnetic fields ( and the velocity and magnetic vector potential fields by extension) appeared to hold true when $\mathrm{Pr_M}=1$ (where $\nu=\eta=10^{-10}$; weak dissipation). As can be seen in figures \ref{fig:uvsa} and \ref{fig:ovsb}, this is still the case for $\nu=\eta\neq 10^{-10}$ ( of course $\mathrm{Pr_M}=1$). It can also be seen that as $\boldsymbol{\omega}\rightarrow0$, $-\mathbf{B}_\mathrm{rms}\rightarrow0$, indicating a correlation between the growth or decay of the rms strength of the vorticity field and the rms strength of the magnetic field. Although the linear relationship breaks down for the cases where $\mathrm{Pr_M}\neq1$, the aforementioned correlation indeed still holds. This non-linear relationship may be explained simply by noting that by having $\nu\neq\eta$, there is the opportunity that one quantity would grow or decay faster than the other. 

Presented in figures \ref{fig:uvsaallprm} and \ref{fig:ovsballprm} are the cases of $\mathbf{u}_\mathrm{rms}$ versus $-\mathbf{A}_\mathrm{rms}$ and $\boldsymbol{\omega}_\mathrm{rms}$ versus $-\mathbf{B}_\mathrm{rms}$ for all of the simulations that were conducted. Again, it can be seen that as $\mathrm{Pr_M}\rightarrow0$ and $\mathrm{Pr_M}\rightarrow\infty$, the linear relationships seen for the case of $\mathrm{Pr_M}=1$ vanish. It clear that as $\mathrm{Pr_M}$ approaches the bounds at zero and infinity, these non-linear relationships appear to form an "envelope'' around the cases for the various $\mathrm{Pr_M}$. The cases of $\mathrm{Pr_M}=0$ and $\mathrm{Pr_M}\rightarrow\infty$ deserve to be examined in some more detail; we brief comment on them.
 \\
\subsection{The Special Cases of $\mathrm{Pr_M}=0$ and $\mathrm{Pr_M}\rightarrow\infty$}

 \begin{widetext}
~
 \begin{figure}
 	\centering
	 \caption{Simulation results for $\mathbf{u}_\mathrm{rms}$ versus $-\mathbf{A}_\mathrm{rms}$ and $\boldsymbol{\omega}_\mathrm{rms}$ versus $-\mathbf{B}_\mathrm{rms}$ for the case of $\mathrm{Pr_M}=0$.}
         \begin{subfigure}[b]{0.40\textwidth}
        		\includegraphics[width=\textwidth]{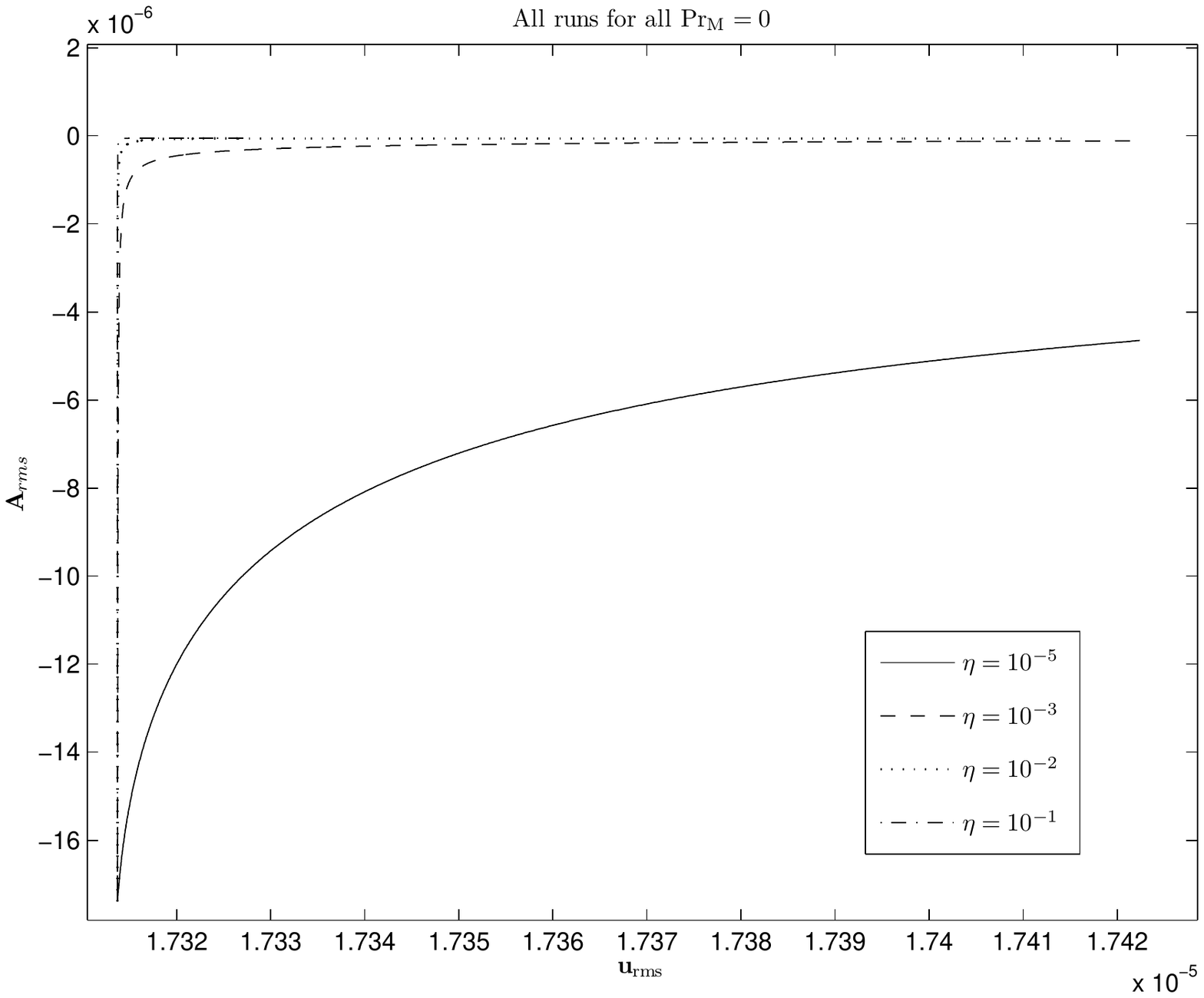}
        			\caption{\it $\mathbf{u}_\mathrm{rms}$ versus $-\mathbf{A}_\mathrm{rms}$ graphically for the case of $\mathrm{Pr_M} = 0$. The curves appear to grow away from the direction of the other curves displayed in figure \ref{fig:uvsaallprm}, implying that the curves for $\mathrm{Pr_M}=0$ do not form an upper ``envelope''.}
     			 \label{fig:uvsaprm0}
    	\end{subfigure}~\qquad\qquad\qquad
	\begin{subfigure}[b]{0.40\textwidth}
       		 \includegraphics[width=\textwidth]{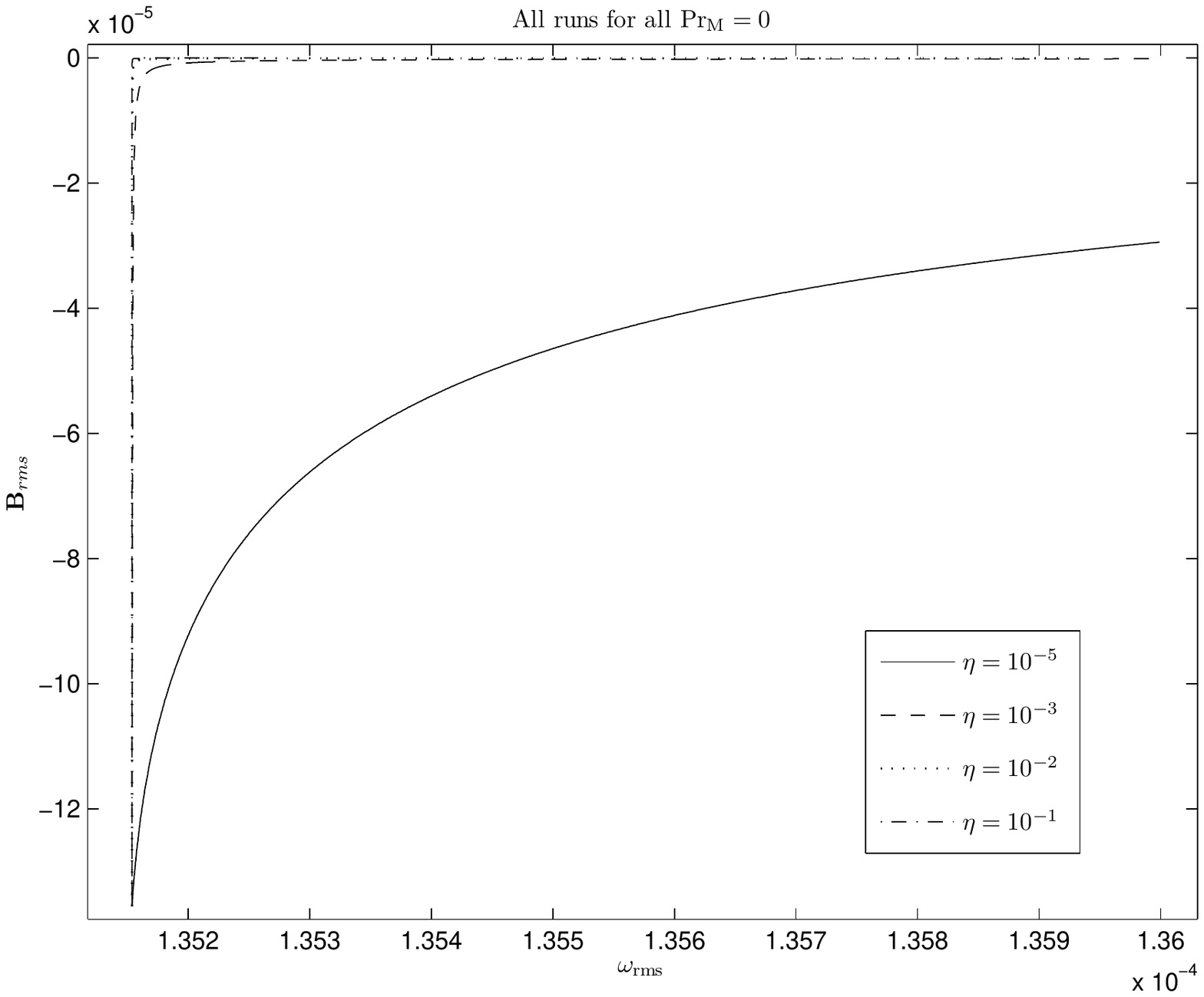}
        		\caption{\it\it $\boldsymbol{\omega}_\mathrm{rms}$ versus $-\mathbf{B}_\mathrm{rms}$ graphically for the case of $\mathrm{Pr_M} = 0$. The curves appear to grow away from the direction of the other curves displayed in figure \ref{fig:ovsballprm}, implying that the curves for $\mathrm{Pr_M}=0$ do not form an upper ``envelope''.}
     			\label{fig:ovsbprm0}
        \end{subfigure}\\
	\centering
	\caption{Simulation results for $\mathbf{u}_\mathrm{rms}$ versus $-\mathbf{A}_\mathrm{rms}$ and $\boldsymbol{\omega}_\mathrm{rms}$ versus $-\mathbf{B}_\mathrm{rms}$ for the case of $\mathrm{Pr_M}\rightarrow\infty$.}
        \begin{subfigure}[b]{0.40\textwidth}
       		 \includegraphics[width=\textwidth]{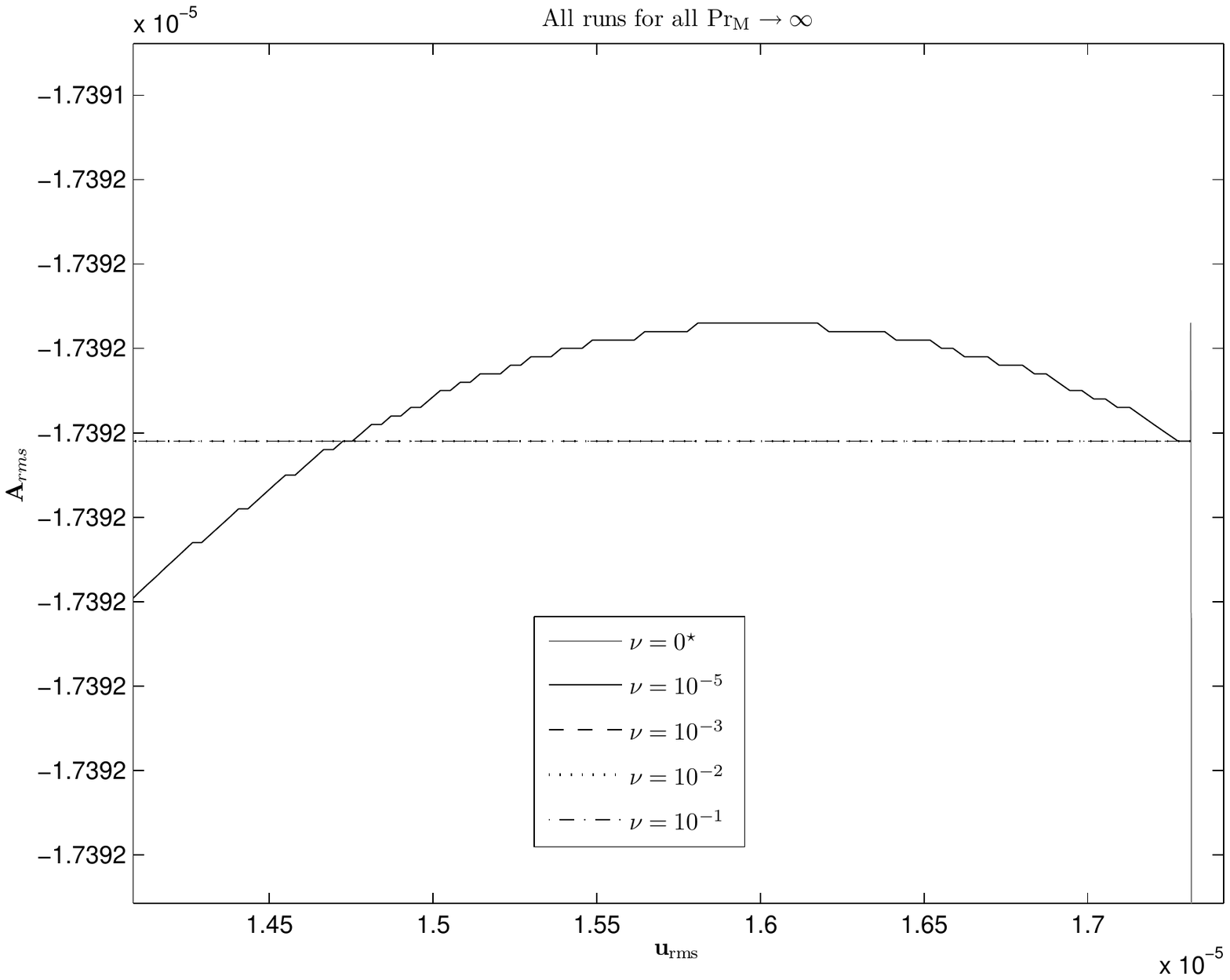}
        		\caption{\it $\mathbf{u}_\mathrm{rms}$ versus $-\mathbf{A}_\mathrm{rms}$ graphically for the case of $\mathrm{Pr_M} = \infty$. The curves appear to grow away from the direction of the other curves displayed in figure \ref{fig:uvsaallprm}, implying that the curves for $\mathrm{Pr_M}\rightarrow\infty$ do not form a lower ``envelope''.}
     			 \label{fig:uvsaprminf}
        \end{subfigure}~\qquad\qquad\qquad
        \begin{subfigure}[b]{0.40\textwidth}
       		 \includegraphics[width=\textwidth]{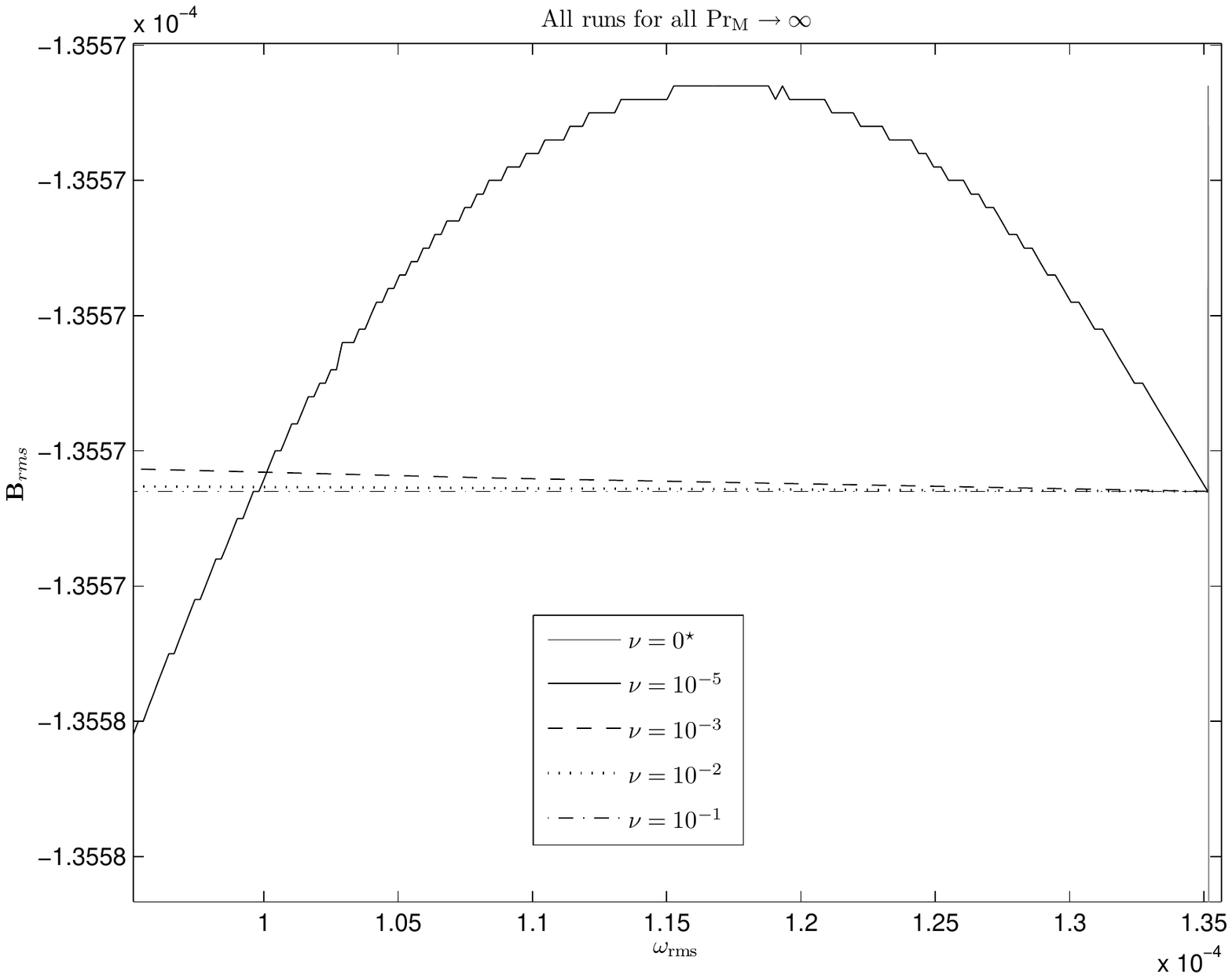}
        		\caption{\it $\boldsymbol{\omega}_\mathrm{rms}$ versus $-\mathbf{B}_\mathrm{rms}$ graphically for the case of $\mathrm{Pr_M} = \infty$. The curves appear to grow away from the direction of the other curves displayed in figure \ref{fig:ovsballprm}, implying that the curves for $\mathrm{Pr_M}\rightarrow\infty$ do not form a lower ``envelope''.}
     			 \label{fig:ovsbprminf}
        \end{subfigure}\\
   \end{figure}
\end{widetext}

In order to investigate the behaviour of the fluids for $\mathrm{Pr_M}=0$ and $\mathrm{Pr_M}\rightarrow\infty$, we present plots of $\boldsymbol{\omega}_\mathrm{rms}$ versus $-\mathbf{B}_\mathrm{rms}$ for these cases in figures \ref{fig:uvsaprm0}, \ref{fig:ovsbprm0}, \ref{fig:uvsaprminf} and \ref{fig:ovsbprminf}.

For the case of $\mathrm{Pr_M}=0$ it is suggested from figures \ref{fig:uvsaprm0} and \ref{fig:ovsbprm0} that $\mathbf{u}_\mathrm{rms}$ versus $-\mathbf{A}_\mathrm{rms}$ and $\boldsymbol{\omega}_\mathrm{rms}$ versus $-\mathbf{B}_\mathrm{rms}$ do not form an upper-enclosing ``envelope'', with the curves for all the simulated values of $\eta$ growing in a direction opposite to what is presented for the case of $0<\mathrm{Pr_M}<1$. This behaviour could possibly be due to systemic effects within the code itself, or could be physically movtivated from the point-of-view of the equations that because of the lack of a diffusive term in eqn (\ref{eq:ns}), the velocity and vorticity field strengths continue to grow very slowly, thus not being able to aid the magnetic vector potential and magnetic field strengths in growing whose equations do have a non-vanishing diffusive term in them. On the side of the magnetic field, diffusion prevails and causes the field and potential strengths to decay overall.

 For the case of $\mathrm{Pr_M}\rightarrow\infty$, an argument similar to the case of $\mathrm{Pr_M}=0$ holds. From figures \ref{fig:uvsaprminf} and \ref{fig:ovsbprminf}, it is suggested that only the case of $\nu=10^{-5}$ will not form part of the lower-enclosing "envelope'' to which the shape of the curves for $1<\mathrm{Pr_M}<\infty$ appear to tend towards, whilst the other cases of non-zero $\nu$ do in fact form part of this ``envelope''. Again, we could attribute the behaviour seen in these plots to systemic effects within the code, but arguing from the point-of-view of the equations once more, we suggest that even these other cases of non-zero $\nu$ will not form part of this ``envelope''. This is due to the fact that whilst both of the Induction Equations do not possess a diffusive term, allowing for a slowly-growing magnetic field, the Navier-Stokes and Vorticity equations do, which causes the velocity and vorticity rms strengths to decay at a faster rate than that of the growth of the magnetic field and magnetic vector potential rms strengths. We suggest then that over time, the magnetic field and magnetic vector potential rms strengths would also begin decaying due to the decaying contribution of the velocity field strength in the induction term of the Induction Equations. Once more, this would cause the curves for the case of $\mathrm{Pr_M}\rightarrow\infty$ to grow in a different direction to that of the other curves and thus not form part of the ``envelope''. 

It should be noted in figures \ref{fig:uvsaprminf} and \ref{fig:ovsbprminf} that the case for $\nu=0$ is special, as it represents a magnetic Prandtl number of $0/0$, which is undefined. This case is itself degenerate and is only included for completeness.

Even though we have provided a physical explanation from the point-of-view of the equations for the results obtained above for the special cases of $\mathrm{Pr_M}=0$ and $\mathrm{Pr_M}\rightarrow\infty$, further study is still needed. Physically, a neutral fluid having zero viscosity is often referred to as a superfluid. These superfluids are known for not behaving in the standard ways described by Fluid Dynamics, meaning that care should be taken when simulating them numerically. A similar argument can be made for charged fluids with zero and infinite magnetic diffusion. We thus conclude that the results obtained for these cases should be treated with utmost care and not readily be attributed to any physical behaviour.

\section{Conclusion}

In this paper we have presented simulations exploring the analogy between vorticity and magnetic fields, with a strong focus on technical details, and have found strong evidence in support of an analogy between the two fields. The results in this article (i) reaffirms what is known from previous analysis and (ii) extends the analysis to a full spectrum of fluids. It is clear that both the evolution equations and the corresponding simulations indicate that the two fluids are analogues.  For simplicity, we had neglected the effects of any "battery terms'' in the relevant equations, as well as the Lorentz Force term in the Navier-Stokes equations, deferring these considerations to later works. We also investigated the conditions under which we could achieve exponential growth in the rms field strengths of the velocity, vorticity, magnetic vector potential and magnetic flux density fields and found that when $\nu=0$, exponential growth was achieved for the velocity and vorticity fields' rms strengths, and when $\mathrm{Pr_M}\rightarrow\infty$, exponential growth was achieved for the magnetic vector potential and flux density fields' rms strengths. As expected, the magnetic Prandtl number ($\mathrm{Pr_M}$) played no role in seeking exponentially-growing strengths for the velocity and vorticity fields due to the absence of the Lorentz Force term in the Navier-Stokes equations. Further investigation into the exact nature of $\mathrm{Pr_M}$ and the conditions under which these fields would achieve exponential growth is needed in order to confirm any definite relationship. In addition, further investigation into the nature of the relationship between $\mathrm{Pr_M}$ and the analogy between the vorticity and magnetic fields in the presence of a battery term is also needed and will be investigated in a later work.

\section{Acknowledgements}

Patrick W. M. Adams acknowledges funding support from the National Research Foundation (NRF) of South Africa, as well as funding from the University of Cape Town, both administered by the Postgraduate Funding Office (PGFO) of the University of Cape Town. Bob Osano acknowledges URC funding support administered by the University of Cape Town. A special vote of thanks also goes to the members of the \texttt{pencil-code-discuss} group for technical advice on the \textsc{Pencil Code}.

\bibliography{references} 

\appendix

\end{document}